\documentclass[12pt]{article}

\usepackage{mathrsfs,amsbsy,amssymb,latexsym,amsfonts,amsmath}
\usepackage[dvips]{graphicx,color}

\global\arraycolsep=3pt
\newcommand\blue[1]{\textcolor{blue}{#1}}

\newcommand\lnf{ln \, f}

\newcommand\bref[1]{(\ref{#1})}

\usepackage{amsbsy}
\usepackage{alltt}
\begin{document}
\begin{flushright}
OIQP-09-08 \\
RIKEN-TH-162
\end{flushright}

\vspace*{0.5cm}

\begin{center}
{\Large \bf 
Card game restriction in LHC can only be successful!
} \\
\end{center}
\vspace{10mm}

\begin{center}
{\large
Holger B. Nielsen}\\
{\it
The Niels Bohr Institute,
University of Copenhagen,
\\
Copenhagen $\phi$, DK2100, Denmark
}
\\
and
\\
{\large
Masao Ninomiya} \\
{\it
Okayama Institute for Quantum Physics,\\ 
Kyoyama 1, Okayama 700-0015, Japan \\
{\footnotesize and} \\
Theoretical Physics Laboratory,\\ 
The Institute of Physics and Chemical Research (RIKEN),\\ 
Wako, Saitama 351-0198, Japan.}
Keywords: {Proposal of experiment, Risk in LHC, Higgs particle, \\
Backward causation \qquad \qquad \qquad \qquad \quad \quad}
\end{center}

\vspace{1cm}

\begin{abstract}

We argue that a restriction determined by a drawn card or
quantum random numbers, on the running of LHC (Large Hadron Collider), 
which was proposed in earlier articles by us, 
can only result in an, at first, apparent success whatever the outcome.
This previous work was concerned with looking for backward causation and\,/\,or
 influence from the future, which, in our previous model, was assumed 
to have the effect of arranging bad luck for large Higgs producing 
machines, such as LHC and the never finished 
SSC (Superconducting Super Collider) stopped by Congress because of such bad luck, 
so as not to allow them to work.

\end{abstract}

\newpage
\tableofcontents
\vspace*{0.5cm}

\section{Introduction}

~~~~In the previous articles \cite{search} we proposed that one should use 
the LHC-machine to look for 
backward causation effects. 
Indeed, we proposed a model \cite{search,3,4,5} 
in which the realized history of the 
universe was selected so as to minimize a certain functional 
of the history, 
a functional being the imaginary part of the action $S_I$\,[history], 
which only exists in our model.
In general, it is assumed in science that there is no pre-arrangement \cite{r1} 
of initial conditions so as to make special events occur or 
not occur later.
However J.~Bell proposed BBC as a solution to the problems 
of Einstein - Podlosky - Rosen's ``super-determinism'' \cite{6}. 
Also, one of the present authors (H.B.N.) and his group 
earlier proposed models nonlocal in time (and space) \cite{7,8,9}.
Similar backward causation effects have also been proposed 
in connection with the story that e. g. humanity would cause 
a new vacuum to appear, ``vacuum bomb,'' 
by one of the present authors (H.B.N.) and collaborators \cite{10}. 
Our proposal is to test if there should perhaps be such 
pre-arrangements in nature, that is, pre-arrangements that prevent Higgs 
particle producing machines, such as LHC and SSC, from being functional. 
Our model with an imaginary part of the action \cite{search,11}
begins with a series of not completely convincing, 
but still suggestive, assumptions that lead to the 
prediction that large Higgs producing machines should turn out not 
to work in that history of the universe, which is actually 
being realized.

The plan behind the practical experiment, which we proposed, 
was to produce some random 
numbers--partly by drawing cards and partly by quantum random number 
generation -- and then let these random numbers be translated, according 
to the rules of the game, into some restrictions on the luminosity or 
the energy or both of the LHC. Thus LHC might, for instance, only be 
allowed to run up to a certain beam energy.
I.~Stewart \cite{11} proposed that pauses are determined by random numbers.

The idea is merely to require any restriction at all for LHC  
with probability $p$ that is deemed, by the rules of the game, 
to be very small. 
The probability $p$ for a ``close LHC'' card is \,
$p\sim 10^{-6}$ or so.

It is clear that even a small probability restriction being enforced on LHC, 
its luminosity or beam energy, 
means an artificially imposed -- one would say, ignoring our 
type of model, unnecessary -- 
risk for the LHC project.

It is, 
however, 
the main focus of the present letter to point out 
(as was briefly started in the previous article \cite{search}) 
that 
even though our proposed project of restricting LHC 
according to random numbers seems to give rise to a loss, 
in fact, whatever happens seems 
-- initially at least -- to be a gain, a success!

That a success in this sense is guaranteed to be the result 
seemingly with almost 100\% certainty (but in reality not quite 100\%) 
is demonstrated in the present article.

\section{Card game for LHC restrictions can only be a success!}

~~~~There are two possibilities.
\begin{enumerate}
\item[1)] You draw a card combination of the most common type leading to no restrictions.
      Then LHC can run without any restriction and you can be totally happy 
      because you found, with close to zero expense, an argument 
      against our theory.
      You almost kill our theory, 
      or at least drastically diminish the chance that it is right. 
      This is a very good scenario!
\item[2)] You draw a restriction card combination. 
      Now, it is a significant loss that LHC cannot run in full, but
      now you have proved our, or a similar, backward causation theory.
      This would be so interesting, if one really had backward causation, 
      that it might be counted as a discovery greater than supersymmetric partners 
      or the finding of the Higgs.
      It would be a fantastic discovery made with LHC! 
      If the restriction drawn is not a totally closing, you would likely soon also 
      find the Higgs and perhaps the supersymmetric partners 
      even if 
      statistics might initially be a bit worse than hoped for. 
\end{enumerate}

 It would be a wonderful victory for CERN and LHC to find 
\underline{backward causation} 
together with having to obey the most likely very mild restrictions.
 We should remember that the rule of our card game should be 
to make the milder restriction have a much higher chance 
of being drawn than the very strong restriction of, for example, totally closing LHC.

Quite correctly, there is, though little chance of, 
a true loss even though it will not be initially noticed. 
It is possible, although not likely, 
that a random number game leads to a restriction even if our model, 
and any model with backward causation, is wrong.
In this case, we have a bad bargain:
not only would we loose the full applicability of LHC, 
but we would also have gotten, by a statistical fluctuation, the wrong impression 
that a backward causation containing model were indeed true without 
this actually being the case. 

We should certainly arrange the restriction probability $p$ to be low 
enough to make this bad case have a very low probability.

One would, from this way of arguing, initially suspect 
that it would be most profitable not to perform our random number 
LHC restricting experiment 
because
if our theory were right LHC would, in any case, be closed or restricted somehow 
by prearranged bad luck,
as happened to SSC, 
for which Congress in the U.S.A. terminated economic support.
Now, 
however, 
we want to argue that it would be more agreeable to have LHC be stopped 
or restricted by a random number game rather than by some bad luck 
such as political withdrawal of support. 
The main reason for the artificially caused random number withdrawal being 
preferred is that we would, in this case, get more solid support 
for our, or a similar, model being true than by the same restriction 
coming about through a bad luck accident.

To see that would be more convincingly shown 
the truth of our theory of imaginary action determined 
by history  
if we have a card or random number closure rather than a ``normal'' failure, 
we could contemplate
how much more convincing our theory would have been today 
if the SSC-machine had been closed after a random number experiment 
rather than mainly for economical reasons or perhaps because of 
the collapse of the Soviet Union, 
which made the competition 
with 60 million dollar accelerators not worthwhile. 

Now it is sometimes explained that SSC \cite{12} 
had bad luck because of various stupidities or accidents, 
but had it been a card game nobody could come up with such foolish cards. 
Everything is an accident, 
but we would know the probabilities very reliably. 
So if the card game were set up so that the closing probability 
were sufficiently small, we would have been sure that the closing of 
SSC were due to a (anti)miracle. 

In the following, we shall present a little calculational example to 
illustrate formally that a more reliable knowledge of the truth of 
our theory is obtained with a random number experiment. 
This comes under the discussion of point 2) among the reasons for conducting 
our proposed experiment later in the present article.

\section{Reasons for conducting our proposed experiment}

~~~~What could be a reason to conduct the card game experiment? 
\begin{enumerate}
\item[1)] To obtain greater conviction about the truth of our theory  \\
   -- if it is true of course. -- 
\item[2)] To perhaps avoid bad backward causation effects. 
\end{enumerate} 
These are the two benefits you could have.

In formula it would mean that we should estimate averages 
for the two measures of these two benefits. 

\subsection{3.1-a~~More conviction of truth of our model} 

~~~~For reason 1) -- the conviction about our theory 
that it is indeed right -- we need some measure. 
Both the result of the card game and the failure of the LHC 
for other reasons are statistical events, 
but, while we have very trustable ideas about what probability 
$p$ 
to assign to a given class of card combinations, 
our assignment of a trustable value for the failure probability 
$f$ 
for other reasons is very difficult and has a huge uncertainly. 
Therefore, if LHC fails for a reason other than a random number game, 
we would have not even truly learned that our theory was right 
even though we would say 
``it is remarkable that the present authors wrote about the failure 
while LHC still looked to be able to work.'' 

\subsection{3.1-b~~Miraculocity and estimating evidence for our model}

~~~~In order to understand why the difference between getting 
our model supported by a ``natural '' failure of LHC and a failure 
caused merely by having a card game drawing a 
``restrict LHC'' card gives rise to an important difference in trustability 
in our model. 
 We shall give a slight formal illustration using the statistical model 
which is not very exact but is appropriate for illustrating our point. 

 If, in our model, a seemingly other reason failure of LHC 
occurs merely through the coincidence of a series of small bad luck events  
-- that by themselves can easily happen -- 
then the number and unlikeliness of elements in this series of 
bad luck events must be proportional to 
$  - \lnf = | \lnf| $
where
$f$
is the probability of failure. 
We could call this quantity
$ - \lnf $ 
the ``miraculocity'' for failure in a seemingly natural way. 
This concept of 
``miraculocity''
becomes a measure for how many 
``submiracles''
must occur. 
As examples of submiracles, there are  
``the watch man having drunk a bit too much'', 
``the connection between super conducting cables 
  having too high resistance'', 
``The accident being in the difficult part of the tunnel, 
  just under Jura mountain'' etc.

Now if we set up a card or quantum mechanically based 
random number generator leading to 
``restrict LHC''
with probability  $p$ , 
it needs to generate 
-- by the selection of the realized history in our model -- 
a number of adjusted accidents (or submiracles) 
in a number proportional to 
$ -ln\, p =|ln\, p| $ .
Essentially, in the case of the truth of our theory, whether the failure 
of the LHC will arise via the card or the quantum random number game 
or via a natural reason will depend on which of the two alternative 
miraculocities 
$ -ln\, p $
or 
$ -ln\, f $
is the smallest. 
There will, of course,  be a preference with ``miraculocity'': 
the least miraculous of the two alternative possibilities for failure will 
most likely be the one that occurs. 
This would require fewest submiracles. 

We can define $f$ so that indeed 
$-ln\, f$ 
gives a measure of the ``miraculocity'', but it is very difficult 
even for people building the LHC, to convincingly figure out 
what to accept or predict about this miraculocity 
$ -ln\, f =|ln\, f| $ .
At best, one can predict it with an appreciable uncertainty. 
That is to say, we obtain, at least from some simulation 
-- say by Monte Carlo methods or  
just theoretically -- a probability distribution for 
``miraculocity'' $|ln\, f|$. 
 To illustrate our point of estimating the degree of 
conviction, which we shall obtain in the case of a 
``natural'' and\,/\,or ``normal'' 
failure, we can assume that the probability calculation -- by (computer) 
simulation of the political and technical procedures around 
CERN and LHC -- led to a Gaussian distribution for the miraculocity 
$-ln\, f$.
That is to say, we assume the probability distribution
  \begin{eqnarray}
     && P\left(| ln\, f | \right) d\,|ln\, f|  =  \nonumber \\ 
     && \approx \frac{1}{\sigma \sqrt{2\pi  }}\> \mathrm{exp} 
        \Big( \frac{1}{2\sigma^2}\left(|ln \, f|-\ |ln \, f_0 |\right)^2 
        \Big) \ d\,|ln\, f|.\nonumber \\ 
      \label{Gauss}
  \end{eqnarray}
Here, $\sigma$ is the spread of the distribution for the logarithm of 
$f$ , i.e., 
the ``miraculocity.'' 

Now let us consider the degree of remarkableness for the failure 
depending on whether it is due to the card 
or the quantum random number game 
or a ``normal'' failure, i.e., other reasons 
such as meteors and bad electrical connection between the superconductors. 

In the case of a card or quantum random number game, the number of 
sub-miracles in the card or quantum packing is proportional to 
$-ln\,p$, 
where $p$ is the arranged probability by the game rules. 

However, if there is instead a ``normal'' failure due to the stupidity of some 
members of cabinet or the like, then we would tend, of course, 
to believe that the true miraculocity 
$-ln\, f = |ln\,f|$
for that failure is indeed in the low end of the estimated Gaussian 
distribution.
In other words, we would expect that, after all, the ``true'' probability 
for failure $f$ is rather high, i.e., 
$f>f_0$ 
or presumably even 
$f \gg f_0 $ 

Let us indeed evaluate the expected probability for a seemingly 
``normal'' (i.e., not caused by card etc games) failure. 
This expected normal probability for failure is 
\begin{eqnarray}
\left\langle f \right\rangle
  =\int ^{\infty }_{-\infty }\frac{1}{\sigma \sqrt{2\pi }}\cdot f 
   \cdot \mathrm{exp} \left(-\frac{1}{2\sigma ^2}(ln\, f-ln\, f_0)^2 \right)
d\,|ln\, f|
\end{eqnarray}
(we imagine that the miscalculation by including the 
$f>1$ region is negligible, but one could of course do better if needed). 

  We immediately write 
$f=e^{-|ln\,f|}$. 
We had hoped to expect ``normal'' failure 
with the probability given by 
\begin{eqnarray}
\left\langle f \right\rangle 
    &&=\int ^{\infty }_{-\infty }\frac{1}{\sigma \sqrt{2\pi }}
       \cdot \mathrm{exp} \left(-\frac{1}{2\sigma ^2}(|ln\, f|-|ln\, f_0|)^2 
       -|ln\, f| \right)d\,|ln\, f| \nonumber\\
    &&=\int ^{\infty }_{-\infty }\frac{1}{\sigma \sqrt{2\pi }} \cdot \mathrm{exp} 
       \biggl(-\frac{1}{2\sigma ^2}\left[ \left(|ln\, f|-|ln\, f_0|
       +\sigma ^2\right)^2  \right. \nonumber \\
    && \qquad \qquad \qquad \qquad \qquad \qquad \qquad \quad      
         \Bigl. -\sigma^4 +2\sigma^2 |ln\,f_o| \Bigr] \biggr) d\,|ln\, f| \nonumber\\
    &&=\mathrm{exp}\left( \frac{\sigma^2}{2}-|ln\, f_o|\right) =f_o\,e^{\sigma^2 /2}. 
\end{eqnarray}

 Hence the remarkability or apparent miraculousness of the 
outcome that LHC should fail seemingly by a ``normal'' accident 
-- such as political closure -- 
is not the ``miraculocity'' corresponding to the most likely value for 
$f$, i.e., $-ln\,f_o =|ln\, f_o |$, but rather to 
$\mathrm{``remarkableness''}=-ln\left\langle f \right\rangle =|ln\left\langle f \right\rangle\!|=|ln\,f_o|-\frac{\sigma^2}{2}$. 

It is this correction by the term 
$-\frac{\sigma^2}{2}$ 
that causes less conviction for our model being true 
if the failure of LHC shows up as a ``normal'' failure, 
than if we get a failure caused by a card or quantum random number game. 
One should keep in mind that whether in our model one or the other 
reasons for failure occurs depends largely on the relative sizes of 
$-ln\, f$ and $-ln\, p$ . 

 In this way, it would be more convincing that 
our theory were true if the failure were found by a card game or the like than by a 
``normal'' failure of LHC.
 It would thus be profitable scientifically if we could provoke a 
card game failure instead of a ``normal'' one; we would have 
the possibility of arranging that if our model were right. 
 In the case of our model being wrong, of course, the card game project 
would only add to the totally failure probability of LHC, 
making a card game a risk and a bad thing.

 Should our theory be right, the failure of LHC would be guaranteed 
with $\frac{2}{3}$ probability, and in that case, the chance of total 
failure probability would not change greatly whether we perform 
a card game project or not.
 In that case we would just move some failure probability from the 
``normal'' failure due to the card game or the similar case.

If we place some economical value on the degree of confidence 
we would obtain if our model were indeed true depending on 
whether one failure or another really occurred, 
we could put this benefit into the form 
\begin{eqnarray}
  b_{1)}&&=c\cdot \mathrm{``remarkableness''} \nonumber \\
        &&=c\cdot
        \left\{ \begin{array}{l}
           |ln\,p| \ \mathrm{if\ game\ failure} \\
           |ln\left\langle f \right\rangle\!|=|ln\,f_o|-\frac{\sigma^2}{2}
           \ \mathrm{if\ ``normal''\ failure}. 
        \end{array} \right.  
\end{eqnarray}

In the case of our theory being right, which occurs with probability 
$r$, we estimated that LHC would be stopped with $\frac{2}{3}$
probability \cite{search} so that this benefit will be calculated as an average, 
\begin{eqnarray}
  \left\langle b_{1)}\right\rangle 
     && =c\cdot \mathrm{``remarkableness''} \nonumber \\
     && =c \left\langle \Biggl( \frac{p}{f+p} |ln\,p|+\frac{f}{f+p}
         \left(|ln\, f_o |- \frac{\sigma^2}{2} \right) \Biggr) r\frac{2}{3}
         \right\rangle_{\mathrm{Gauss}},
\end{eqnarray}
where the average $\left\langle \cdots \right\rangle_{\mathrm{Gauss}}$ 
is merely the average over distribution \bref{Gauss} .

 For instance, in the limit of a very small probability $p$ assigned to 
the random number restricting LHC, we would get 
\begin{eqnarray}
\left\langle b_{1)}\right\rangle \approx 
    c\left(|ln\, f_o |- \frac{\sigma^2}{2} \right) r\cdot\frac{2}{3}
    +cp\left\langle \frac{1}{f} \right\rangle \left(|ln\,p|-|ln\,f_o | +
    \frac{\sigma^2}{2}\right) r\> \frac{2}{3}+ \ldots . \label{lowp}
\end{eqnarray}
If, on the other hand, we set $p\gg \left\langle f \right\rangle$, 
we would get
\begin{eqnarray}
\left\langle b_{1)}\right\rangle \approx  \Biggl(|ln\,p|+
\frac{\left\langle f \right\rangle}{p} \left(|ln\,f_o |- \frac{\sigma^2}{2}
-|ln\,p|\right) \Biggr) r\cdot \frac{2}{3}. 
\end{eqnarray}

It is important to notice that, as the previous discussion suggested, 
the correction term in \bref{lowp}
will, for small enough $p$, give increasing benefit with increasing $p$ 
so that it would be beneficial $w. r. t.$ this benefit 
$b_{1)}$ of attaining an increase in the safety of our knowledge 
that $p$ is not completely zero in our model.

\subsection{3.2~~~~~Avoiding bad backwardly caused events}

~~~~In our earlier paper, 
we included, in our estimates of whether it would pay to perform our card game 
or random number game experiment, the consideration 
that if we indeed have backward causation for LHC becoming inoperable, 
then these pre-arrangements could have side effects that might be bad 
and, a priori, perhaps also good.
 The backward causation effects might end up being huge in much the same way 
as the famous forward causation effect of the butterfly in the 
``butterfly effect'', but in the same way as it is difficult to predict 
whether the effects are good or bad when the butterfly beats its wings 
in a particular way, 
it is hard to know if the pre-arrangements set up to prevent LHC  
from working are good or bad.
 If we think of such possibilities as the closure of CERN or an earthquake 
in Geneva, 
we may judge it to be bad, but if we think of even earlier or further distant 
pre-arrangements, it becomes increasingly difficult to estimate 
either good or bad.
For instance, it is a possibility that a major factor behind the SSC
being terminated by Congress was 
the collapse of the Soviet Union \cite{13}. 
 This were a huge backward causation effect but it is hard to evaluate the probability 
as to whether it is good or bad. 
 Thus, it would have been hard to evaluate, in advance, whether 
our card game would have been profitable had our theory been known then.

In the previous articles \cite{search}, 
we called the price of the damage arising in excess 
when a ``normal'' failure of LHC is provoked, $d$ . 

We should imagine that the very huge backward causation effects occurring 
very remotely  
from the LHC are probably averaged out to zero, 
similar to the far future effects of the butterfly wing. 
 Hence the important contributions to the damage cost 
 $d$ are rather close in time (and space) to the LHC itself. 
 We very roughly estimated, in our previous study, 
$d \approx 10 \cdot $ ``cost of LHC'' $\simeq 10 \cdot 3.3 \cdot 10^9$ CHF
$=3.3 \cdot 10^{10}$ CHF. 

 In the case of the card game failure, there may also be huge effects, 
but now the evaluation of the damage being good or bad would be totally 
opaque.
 Only the effects of performing the actual experiment may have any predictable 
average effect. Therefore, in the case of such an artificial failure, 
the damage would be limited \blue{to} statistically washing out damage 
(i.e., they are equally likely to be good or bad) and the obvious loss 
because of the  restriction on the  
$d_\mathrm{\,rest.loss}$ card drawn. 

We should arrange the latter damage to almost certainly be the minimal 
one by assigning mild restrictions to be much more likely outcomes than 
heavy restrictions. 

The damage done, or by switching the sign, the (negative) benefit, is 
\begin{eqnarray}
-b_{2)}=d\cdot\frac{2}{3}r\cdot \frac{f}{f+p}+d_\mathrm{\,rest.loss} \cdot
\Biggl( p\left(1-\frac{2}{3}r\right)+\frac{2}{3}r\cdot \frac{p}{f+p}\Biggr) , 
\end{eqnarray}
where we used the notation $d_\mathrm{\,rest.loss}$ for the cost of the restrictions. 

\section{Conclusion}

~~~~We have argued that it would be a good idea to perform 
our earlier proposed experiment of generating some random numbers 
-- by card drawing or by a quantum random number generator, 
or even both ways -- 
and letting them then be decisive in applying restrictions on the beam energy and/or 
the luminosity and/or the like. 

The main point was that our theory, referred to as 
``model with an imaginary part of the action'', 
is indeed right if LHC is 
stopped by our proposed game than if it just failed for some technical 
or political reason. 
The reason for it being more convincing that the stoppage of LHC is due to 
a random number rather than a ``normal'' failure is that it is very hard to estimate 
in advance how likely it is for a ``normal'' failure of LHC to occur. 

The greatest encouragement for performing the experiment without much hesitation 
is the remark that  
whatever happens with our proposed experiment, it will, in practice, seem to 
be a success or at least to be of no harm. 
The point is that in the case of any restriction being imposed 
by the random numbers, we have, because of the very fact of 
these random numbers being generated at all, obtained the shocking 
great discovery that there is ``backward causation.'' 
Such a discovery of the future influencing the present and past would 
be monumental. 
Consequently, we would be very happy and it would be a fantastic success 
for the LHC to have caused such a discovery!
\renewcommand{\thesection}{}
\section{Acknowledgments}

~~~~The authors acknowledge the Niels Bohr Institute and 
Okayama Institute for Quantum Physics for the hospitality 
extended to them. This work was supported by the Grants -- 
in-Aids for Scientific Research, No. 19540324,
and No. 21540290, from the 
Ministry of Education, Culture, Sports, Science and Technology, Japan. 


\end{document}